\documentclass[a4paper,11pt]{article}
\usepackage{jheppub} 
\usepackage{lineno}
\usepackage{mathrsfs}

\def\be{\begin{equation}}
\def\ee{\end{equation}}
\newcommand{\scri}{\mathscr{I}}
\newcommand{\scp}{\scri^+}
\newcommand{\bes}{\begin{subequations}}
\newcommand{\ees}{\end{subequations}}
\newcommand{\bea}{\begin{eqnarray}}
\newcommand{\eea}{\end{eqnarray}}

\newcommand{\ricci}{\mathcal R}

\usepackage{xcolor}
\usepackage[normalem]{ulem}

\title{Fully nonlinear transformations of the Weyl-Bondi-Metzner-Sachs asymptotic symmetry group}

\newcommand{\CornellPhys}{Department of Physics, Cornell University, Ithaca, New York 14853, USA}

\newcommand{\CornellCCAPS}{Cornell Center for Astrophysics and Planetary Science (CCAPS), Cornell University, Ithaca, New York 14853, USA}

\newcommand{\UVA}{Department of Physics, University of Virginia, P.O.~Box 400714, Charlottesville, Virginia 22904-7414, USA}

\author[a,b]{\'Eanna \'E. Flanagan}
 \affiliation[a]{\CornellPhys}%
 \affiliation[b]{\CornellCCAPS}%

\author[c]{David A.\ Nichols}
\affiliation[c]{\UVA}
\date{\today}

\emailAdd{eef3@cornell.edu}
\emailAdd{david.nichols@virginia.edu}

\abstract{The asymptotic symmetry group of general relativity in asymptotically flat spacetimes can be extended from the Bondi-Metzner-Sachs (BMS) group to the generalized BMS (GMBS) group suggested by Campiglia and Laddha, which includes arbitrary diffeomorphisms of the celestial two-sphere.  It can be further extended to the Weyl BMS (BMSW) group suggested by Freidel, Oliveri, Pranzetti and Speziale, which includes general conformal transformations.  We compute the action of fully nonlinear BMSW transformations on 
the leading order Bondi-gauge metric functions: specifically, the induced metric, Bondi mass aspect, angular momentum aspect, and shear. These results generalize previous linearized results in the BMSW context by Freidel et al., and also nonlinear results in the BMS context by Chen, Wang, Wang and Yau.  The transformation laws will be useful for exploring implications of the BMSW group.}

\begin{document}
\maketitle
\flushbottom

\section{Introduction}
\label{sec:intro}

The systematic study of spacetimes that become asymptotically flat far from an isolated source was initiated in the 1960s.
Bondi, van der Burg, and Metzner~\cite{1962RSPSA.269...21B} as well as Sachs~\cite{1962RSPSA.270..103S} used coordinates (now called Bondi coordinates) that are adapted to the outgoing null rays of an isolated gravitating and radiating system.
Their analysis allowed them to compute the asymptotic Einstein equations in Bondi coordinates and the symmetry transformations that preserve the metric at infinity and the Bondi coordinate conditions.
The group of such transformations, the Bondi-Metzner-Sachs (BMS) group, is larger than the Poincar\'e group of transformations that preserve Minkowski spacetime.
Lorentz transformations are contained in the BMS group, and the group has a
semi-direct product structure similar to that of the Poincar\'e group.
However, the infinite-dimensional commutative group of supertranslations replaces the four-dimensional group of spacetime translations in the Poincar\'e group, though the spacetime translations remain a subgroup of the supertranslations.

More recently, the analysis of asymptotically flat spacetimes has been revisited, and larger groups of symmetry transformations were found which preserve different geometric quantities at future null infinity (this is described in more detail in Sec.~\ref{sec:review})
\cite{Stro-lectures,2023arXiv231012922D}.
These new groups include the extended BMS group~\cite{Barnich:2009se,Barnich:2010eb}, the generalized BMS group~\cite{Campiglia:2014yka}, and the Weyl-BMS (BMSW) group~\cite{Freidel:2021fxf,Freidel:2021qpz,Freidel:2021dfs,Freidel:2021ytz,Chandrasekaran:2021vyu}. 
While the action of BMS transformations on the Bondi metric functions
has been computed to nonlinear order \cite{2002hftr.book.....C,FN,Flanagan:2021svq,Chen:2023zpl},
the actions of the extended BMS, generalized BMS and BMSW transformations have
been computed to linear order only \cite{Compere:2018ylh,Freidel:2021fxf}.
In this paper, we derive the nonlinear transformations of the leading-order metric functions in vacuum for the generalized BMS and BMSW groups.
These transformations reduce to the known nonlinear BMS and linearized BMSW results in the appropriate limits.

Fully nonlinear transformation laws can be useful for exploring the viability of alternative
  definitions of charges associated with symmetries.  An example is 
  the recent investigation of Ref.\ \cite{Chen:2022fbu} of continuity
  of various charges as the cross-section of future null infinity is
  varied.  They are also useful for understanding the space of vacua
  which is relevant to the quantum theory \cite{Compere:2018ylh,Freidel:2021fxf}.

The remainder of this paper contains a review of the different
asymptotic symmetry groups in Sec.~\ref{sec:review}.
The main results are derived in Sec.~\ref{sec:transformations} and compared with existing results in Sec.~\ref{sec:comparison}.
Some applications of these results are given in Sec.~\ref{sec:applications}.

\section{Review of asymptotic symmetry groups in asymptotically flat
  spacetimes}  \label{sec:review}

In this section, we review the different asymptotic symmetry groups
in asymptotically flat spacetime to establish our notation and conventions. 

We use retarded Bondi coordinates $(u,r,\theta^1,\theta^2) =
(u,r,\theta^A)$ near future null infinity, following
Refs.\ \cite{1962RSPSA.269...21B,Barnich:2009se,Barnich:2010eb,Strominger:2013jfa,He:2014laa,Kapec:2014opa,Strominger:2014pwa,Pasterski:2015tva,FN,Compere:2018ylh, 
  Freidel:2021fxf,Freidel:2021qpz}.  We will use throughout the
notations and conventions of Flanagan and Nichols~\cite{FN}
(henceforth FN).  The key metric functions in Bondi gauge are the
induced metric $h_{AB}(\theta^A)$, the Bondi mass aspect
$m(u,\theta^A)$, the angular momentum aspect $N_A(u,\theta^A)$, and the
shear $C_{AB}(u,\theta^A)$.
The metric expansion obtained from the asymptotic conditions, the gauge conditions and the vacuum Einstein equations is
\begin{eqnarray}
  ds^2 &=& - \left[\frac{1}{2} \ricci - \frac{2 m}{r} + O \left( \frac{1}{r^2} \right)  \right] du^2 
-2 \left[ 1 - \frac{C_{AB} C^{AB}}{16 r^2} + O \left( \frac{1}{r^3} \right) \right] du dr
    \nonumber \\
    &&+ r^2 \left[ h_{AB} + \frac{1}{r} C_{AB}  + \frac{C_{CD} C^{CD}}{4 r^2} h_{AB} +
      \left( \frac{1}{r^3} \right) \right] (d\theta^A -
{\cal U}^A du) (d\theta^B - {\cal U}^B du). \ \ \ \ 
\label{metric}
\end{eqnarray}
Here
\be
    {\cal U}^A = -\frac{D_B C^{AB}}{2 r^2} 
    + \frac{1}{r^3} \bigg[ - \frac{2}{3} N^A
+ \frac{1}{16} D^A(C_{BC} C^{BC})
+ \frac{1}{2} C^{AB} D^C C_{BC} \bigg]
+ O\left(\frac{1}{r^4}\right), \ \ \ \ 
\label{NAdef}
\ee
$\ricci$ is the two dimensional Ricci scalar of
$h_{AB}(\theta^A)$, and $A,B$ are angular indices which run over the values
$1,2$ and are raised and lowered with $h^{AB}$ and $h_{AB}$, respectively.
Also we have generalized the treatment of FN, following
Comp\`ere, Fiorucci and Ruzziconi \cite{Compere:2018ylh}, to allow the
induced metric $h_{AB}$ to differ from the canonical round metric, so that 
$\ricci$ is allowed
to be an arbitrary function of $\theta^A$ instead of being constrained to $\ricci = 2$.
This generalization
requires replacing the leading term in the expansion (FN,2.3b) with
$\ricci/2$, adding the term $D^2 \ricci/8$ to the right
hand side of the evolution equation (FN,2.11a) for the Bondi mass
aspect, and adding the term $C_{AB} D^B \ricci/4$ to the evolution equation
(FN,2.11b) for the angular momentum aspect \cite{Compere:2018ylh}.
In this paper we specialize to spacetimes which are vacuum near
$\scp$, and so the subleading shear tensor ${\cal D}_{AB}$ of (FN,2.3c) vanishes, by (FN,2.10).
  
We will consider three different asymptotic groups
obtained from three different phase space definitions
(see Refs.\ \cite{Chandrasekaran:2021vyu,2023arXiv231012922D} for reviews).
In the coordinate system $y^i = (u,\theta^A)$ on $\scp$,
the diffeomorphisms $\psi : \scp \to \scp$ have the following form for all 
three groups:
\bes
\label{diffeogen}
\bea
\label{barudef}
    {\bar u} &=& e^{\alpha(\theta^A)}  \left [ u + \beta(\theta^A)
      \right],\\
    \label{barthetadef}
    {\bar \theta}^A &=& \chi^A(\theta^B),
\eea
\ees
where $\chi : S^2 \to S^2$ is a diffeomorphism of the two-sphere $S^2$,
and for a point ${\cal P}$ on $\scp$ we have defined $y^i =
y^i({\cal P})$ and ${\bar y}^i = y^i(\psi({\cal P}))$.  
The groups are:
\begin{itemize}
\item The Weyl BMS (BMSW) group suggested by Freidel, Oliveri,
  Pranzetti and Speziale
  \cite{Freidel:2021fxf,Freidel:2021qpz,Freidel:2021dfs,Freidel:2021ytz,Chandrasekaran:2021vyu}.
  For this group the two-sphere diffeomorphism $\chi$ and the
  functions $\beta$ and $\alpha$ can be freely chosen.
    
\item The generalized BMS (GBMS) group suggested by   
  Campiglia and Laddha \cite{Campiglia:2014yka} and further studied in
  Refs.\ \cite{Compere:2018ylh,CL,Donnay:2020guq,Flanagan:2019vbl,Campiglia:2016efb,Compere:2020lrt,Campiglia:2020qvc}.
  For this group the function $\alpha$ is determined as a function of
  $\chi$ as follows.  Let
  $\epsilon_{AB}$ be one of the two volume forms on the two-sphere that are determined 
  up to sign by the metric $h_{AB}$.  Define the function
  $\omega_\chi$ by
\be
\label{not:wdef}
\chi_* \epsilon_{AB} = \frac{1}{\omega_\chi} \epsilon_{AB},
\ee
where $\chi_*$ is the pullback operator.
Then we have 
  \be
  e^{2 \alpha} = \frac{1}{| \omega_\chi|},
  \label{tauval}
  \ee
which will have the consequence that GBMS transformations preserve
  the volume form $\epsilon_{AB}$ up to a sign (see
  Section \ref{2sd:gbms} below).

\item The BMS group \cite{1962RSPSA.269...21B,
    1962RSPSA.270..103S,1962PhRv..128.2851S,Wald:1999wa,Ashtekar:2014zsa},
the subgroup of GBMS for which the diffeomorphisms $\chi$ are
restricted to be global conformal 
  isometries of the two-sphere.  As a consequence the metric $h_{AB}$
  is preserved under BMS transformations.

\end{itemize}

\section{Derivation of transformation laws} \label{sec:transformations}

In this section, we derive the transformation properties of the
metric functions under nonlinear BMSW transformations, for solutions
which are vacuum near ${\mathscr I}^+$.

\subsection{Supertranslations}
\label{sec:sts}

Consider first supertranslations, and specifically the finite supertranslation $\psi :
{\mathscr I}^+ \to {\mathscr I}^+$. Denoting the coordinates
$(u,\theta^A)$ by $y^i$ 
and defining ${\bar y}^i = y^i \circ \psi
 = ({\bar u}, {\bar \theta}^A) $ the mapping $\psi$ on ${\mathscr I}^+$ is given by
\be
\label{st1}
{\bar u} = u + \beta(\theta^A), \ \ \ {\bar \theta}^A = \theta^A.
\ee
Under this mapping the metric functions transform as
\begin{subequations}
\label{nonlinst}
\begin{eqnarray}
  \label{nonlinsta0}
  {\bar h}_{AB}(\theta^A) &=& h_{AB}(\theta^A),  \\
  \label{nonlinsta}
    {\bar C}_{AB}(u,\theta^A) &=& C_{AB}(u + \beta, \theta^A) - 2 D_A D_B \beta
    + h_{AB} D^2 \beta,\\
  \label{nonlinstb}    
    {\bar m}(u,\theta^A) &=& m(u + \beta, \theta^A) + \frac{1}{2} D_A N^{AB}(u+\beta,\theta^A) D_B\beta
    +    \frac{1}{4} N^{AB}(u+\beta,\theta^A) D_A D_B\beta \nonumber \\
    && +  \frac{1}{4} {\dot N}^{AB}(u+\beta,\theta^A) D_A \beta
    D_B\beta + \frac{1}{4} D^A\beta D_A \ricci, \\
      \label{nonlinstc}
    {\bar N}_A(u,\theta^A) &=& N_A(u+\beta,\theta^A) + 3
    m(u+\beta,\theta^A)  D_A\beta
    + \frac{3}{4} D_A \beta D^B \beta D_B \ricci - \frac{3}{8} D_A \ricci (D \beta)^2
    \nonumber \\ &&
    + \frac{3}{4} D_B \beta \left[ D_A D_C C^{BC}(u+\beta,\theta^A) - D^B D^C C_{CA}(u+\beta,\theta^A) \right]
\nonumber \\ &&
+ \frac{3}{4} D_B \beta \, C_{AC}(u+\beta,\theta^A) \, N^{BC}(u+\beta,\theta^A)
+ \frac{3}{2} D_A\beta \,D_B\beta \, D_C N^{BC}(u+\beta,\theta^A)
\nonumber \\ &&
- \frac{3}{4} (D \beta)^2 \,D^C N_{AC}(u+\beta,\theta^A)
+ \frac{1}{2} D_A\beta \, D_B \beta \, D_C \beta {\dot N}^{BC}(u+\beta,\theta^A)
\nonumber \\ &&- \frac{1}{4} (D \beta)^2 \, D^C \beta {\dot N}_{CA}(u+\beta,\theta^A).
  \end{eqnarray}
\end{subequations}
Here dots denote derivatives with respect to $u$, $D_A$ is the
covariant derivative associated with $h_{AB}$, and $\ricci$ is the Ricci
scalar of $h_{AB}$.  Also 
$N_{AB} = {\dot C}_{AB}$ is the Bondi news
tensor, and $(D \beta)^2 = D^A \beta D_A \beta$.
The transformation laws (\ref{nonlinst}) apply to all three groups, BMSW,
GBMS and BMS.

To derive the transformation laws (\ref{nonlinst}), we extend the
diffeomorphism (\ref{st1}) to a diffeomorphism $\eta : M \to M$ which
coincides with $\psi$ on ${\mathscr I}^+$, by
assuming a general power series expansion in $1/r$ and 
using the notation $x^\alpha =
(u,r,\theta^A)$ and ${\bar x}^\alpha = x^\alpha \circ \eta$:
\begin{subequations}
  \label{nonlinst2}
  \begin{eqnarray}
    {\bar u} &=& u + \beta(\theta^A) + \frac{1}{r}
    \beta^{(1)}(u,\theta^A)
    + \frac{1}{r^2} \beta^{(2)}(u,\theta^A) + O\left( \frac{1}{r^3} \right),\\
{\bar r} &=& r + R^{(1)}(u,\theta^A) + \frac{1}{r} R^{(2)}(u,\theta^A)
+ O\left( \frac{1}{r^2} \right),\\
\label{nonlinst2c}
{\bar \theta}^A &=& \theta^A + \frac{1}{r} \chi^{(1)A}(u,\theta^A)
+ \frac{1}{r^2} {\hat \chi}^{(2)A}(u,\theta^A)
+ \frac{1}{r^3} {\hat \chi}^{(3)A}(u,\theta^A)
+ O\left( \frac{1}{r^4} \right).
  \end{eqnarray}
\end{subequations}
Here the functions $\beta^{(1)}$, $\beta^{(2)}$, $R^{(1)}$, $R^{(2)}$,
$\chi^{(1)}$, ${\hat \chi}^{(2)}$ and ${\hat \chi}^{(3)}$ are
arbitrary.
One subtlety is that although $\chi^{(1)A}$ transforms as a vector
under diffeomorphisms of the two-sphere $S^2$, the higher order functions
${\hat \chi}^{(2)A}$ and ${\hat \chi}^{(3)A}$ in Eq.\ (\ref{nonlinst2c}) do not.
We remedy this by parameterizing the diffeomorphism $\varphi : S^2 \to
S^2$ on the two sphere at fixed $u$ and $r$ in terms of three vector fields
$\chi^{(1)A}$, $\chi^{(2)A}$ and $\chi^{(3)A}$:
\be
\label{nonlinst1}
\varphi =
\varphi_{{\vec \chi}^{(3)}}(r^{-3})
  \circ
\varphi_{{\vec \chi}^{(2)}}(r^{-2})
\circ
\varphi_{{\vec \chi}^{(1)}}(r^{-1}) \left[ 1 + O\left(\frac{1}{r^4}
    \right) \right].
\ee
Here for any vector field ${\vec \chi}$ on $S^2$ the map
$\varphi_{\vec \chi}(\varepsilon)$ is the knight diffeomorphism that
moves any point $\varepsilon$ units along the integral curve of
${\vec \chi}$ that passes through that point
\cite{Flanagan:1996gw,Sonego:1997np}. 
By comparing Eqs.\ (\ref{nonlinst2c}) and (\ref{nonlinst1}) we find
that\footnote{This is most easily derived using the formula for the
pullback of a knight diffeomorphism in terms of Lie derivatives: $\varphi_{{\vec
    \chi}}(\varepsilon)_* = \sum_n \varepsilon^n ( \pounds_{\vec \chi})^n
/ n!$.}
\begin{subequations}
    \label{hatchif}
  \begin{eqnarray}
    {\hat \chi}^{(2)A} &=& \chi^{(2)A} + \frac{1}{2} \chi^{(1)B}
    \partial_B \chi^{(1)A}, \\
    {\hat \chi}^{(3)A} &=& \chi^{(3)A} +  \chi^{(1)B}
    \partial_B \chi^{(2)A} + \frac{1}{6} \chi^{(1)B} \partial_B \left(
    \chi^{(1)C} \partial_C \chi^{(1)A} \right).
  \end{eqnarray}
  \end{subequations}

We next take the pullback ${\bar g}_{ab} = \eta_* g_{ab}$ of the metric 
(\ref{metric}) using the expansions (\ref{nonlinst2})
and imposing that it has the same form as
the metric (\ref{metric}), but with
different metric functions which we denote with overbars.
We adopt the shorthand notation that $O(\alpha\beta,n)$ means
the $O(r^{-n})$ piece of the $(\alpha\beta)$ component of this metric
comparison.
The derivation takes place in a series of steps where each step is
simplified using results from the previous steps.
From $O(uu,0)$, $O(ur,1)$, $O(rA,0)$,
$O(rr,2)$ and the trace part of $O(AB,-1)$, respectively, we find
that
\begin{subequations}
  \begin{eqnarray}
    \dot R^{(1)} & =& 0, \\
    \dot \beta^{(1)}& =&0, \\
    \chi^{(1)A} &=& - h^{AB} D_B \beta, \\
    \beta^{(1)} &=& - \frac{1}{2} (D \beta)^2, \\
    R^{(1)} &=& \frac{1}{2} D^2 \beta.
  \end{eqnarray}
  \end{subequations}
We then obtain from $O(AB,-2)$ and from the trace-free part of
$O(AB,-1)$ the transformation 
laws (\ref{nonlinsta0}) and (\ref{nonlinsta})
for the metric $h_{AB}$ and shear tensor $C_{AB}$.

Next we find from $O(rA,1)$, $O(rr,3)$, $O(ru,2)$, and $O(rA,2)$
respectively that
\begin{subequations}
  \begin{eqnarray}
    \chi^{(2)A} & =& \frac{1}{2} C^{AB}(u+\beta,\theta^A) D_B \beta -
    \frac{1}{2} D^A D_B \beta D^B \beta
    + \frac{1}{2} D^A \beta D^2 \beta, \\
    \label{beta2ans}
\beta^{(2)}& =& \frac{1}{4} C_{AB}(u+\beta,\theta^A) D^A \beta D^B
\beta + \frac{1}{4} (D \beta)^2 D^2 \beta, \\
R^{(2)} &=&
- \frac{1}{2} D_A C^{AB}(u + \beta,\theta^A) D_B \beta
- \frac{1}{4} N^{AB}(u+\beta,\theta^A) D_A\beta D_B \beta
+ \frac{1}{4} \ricci
(D \beta)^2
\nonumber \\ &&
- \frac{1}{4} C^{AB}(u+\beta,\theta^A) D_A D_B \beta
+ \frac{1}{4} D_A D_B \beta D^A
D^B \beta - \frac{1}{8} (D^2 \beta)^2, 
\\
\label{chi3}
    \chi^{(3)A} &=& - \frac{1}{16} D^A \beta C_{BC}(u+\beta,\theta^A)
    C^{BC}(u+\beta,\theta^A)
    + \frac{1}{3} N^{AB}(u+\beta,\theta^A) D_B \beta (D \beta)^2 \nonumber \\
&&    - \frac{1}{6} N^{BC}(u+\beta,\theta^A) D^A\beta D_B\beta D_C
    \beta
    + \frac{1}{4} C_{BC}(u+\beta,\theta^A) D^A\beta D^B D^C\beta
    \nonumber \\
    &&    - \frac{1}{2} C_{BC}(u+\beta,\theta^A) D^C\beta D^A D^B\beta
    - \frac{1}{4} D^B C_{BC}(u+\beta,\theta^A) D^A\beta D^C\beta
    \nonumber \\ &&
    + \frac{1}{4} D_B C^{AC}(u+\beta,\theta^A) D^B\beta D_C\beta
    + \frac{1}{2} D^A\beta D_B D^2\beta D^B \beta
    - \frac{1}{3} D^A D_B D_C \beta D^B\beta D^C \beta  \nonumber \\
    && - \frac{5}{24} D^A\beta (D^2 \beta)^2
    + \frac{1}{12} D^A\beta D_B D_C \beta D^B D^C \beta
    + \frac{1}{6} D^A D^B \beta D_B \beta D^2 \beta
    \nonumber \\ &&
    + \frac{1}{6} \ricci D^A\beta (D \beta)^2,
  \end{eqnarray}
\end{subequations}
where $D^2 = D_A D^A$ and $(D \beta)^2 = D^A \beta D_A \beta$.
Finally from $O(uu,1)$ and $O(uA,1)$ we obtain the transformation laws
(\ref{nonlinstb}) and (\ref{nonlinstc}) for the Bondi mass aspect $m$ and angular
momentum aspect $N_A$.

\subsection{Conformal transformations of the Weyl BMS group}
\label{sec:conf}

We now turn to conformal transformations $\psi: {\mathscr I}^+ \to {\mathscr I}^+$ of the form
\be
\label{weyl1}
{\bar u} = e^{\tau(\theta^A)} u, \ \ \ {\bar \theta}^A = \theta^A.
\ee
Under this mapping the metric functions transform as
\begin{subequations}
\label{wnonlinst}
\begin{eqnarray}
  \label{wnonlinsta0}
  {\bar h}_{AB}(\theta^A) &=& e^{-2 \tau} h_{AB}(\theta^A),  \\
  \label{wnonlinsta}
    {\bar C}_{AB}(u,\theta^A) &=& e^{-\tau} C_{AB}(e^\tau u, \theta^A) - 2 u e^{-\tau} \left(D_A D_B 
    - \frac{1}{2} h_{AB} D^2\right) e^\tau,\\
  \label{wnonlinstb}    
        {\bar m}(u,\theta^A) &=& e^{3 \tau} m(e^\tau u, \theta^A)
        +    \frac{1}{4} e^{2 \tau} C^{AB}(e^\tau u,\theta^A) D_A D_Be^\tau
        +    \frac{1}{4} u e^{3 \tau} N^{AB}(e^\tau u,\theta^A) D_A D_Be^\tau
        \nonumber \\
        &&
        + \frac{1}{2} u e^{4 \tau} D_A N^{AB}(e^\tau u,\theta^A) D_B\tau
+ \frac{1}{4} e^{5 \tau} u^2 {\dot N}^{AB}(e^\tau u, \theta^A) D_A\tau
D_B \tau \nonumber \\
&& - \frac{1}{2} u e^{2 \tau} \left[ D_A D_B e^\tau D^A D^B e^\tau -
  \frac{1}{2} (D^2 e^\tau)^2 \right]
        + \frac{1}{4} u e^{4 \tau} D^A\tau D_A \ricci, \\
      \label{wnonlinstc}
    {\bar N}_A(u,\theta^A) &=& e^{2 \tau} N_A(e^\tau u,\theta^A) + 3 u
    e^{3 \tau}  m(e^\tau u,\theta^A)  D_A\tau  \nonumber \\
    &&
+ \frac{3}{4} u^2 e^{4 \tau} \left(D_A \tau D^B \tau D_B \ricci - \frac{1}{2} D_A \ricci
D^B 
\tau D_B \tau \right)
\nonumber \\ &&
+ \frac{1}{2} u^3 e^{5 \tau} D_A\tau \, D_B \tau \, D_C \tau {\dot N}^{BC}(e^\tau u,\theta^A)
- \frac{1}{4} u^3 e^{5 \tau} (D \tau)^2 \, D^C \tau {\dot N}_{CA}(e^\tau u,\theta^A)
\nonumber \\ &&
+ \frac{3}{2} u^2 e^{4 \tau} D_A\tau \,D_B\tau \, D_C N^{BC}(e^\tau u,\theta^A)
- \frac{3}{4} u^2 e^{4 \tau} (D \tau)^2 \,D^C N_{AC}(e^\tau u,\theta^A)
\nonumber \\ &&
    + \frac{3}{4} u e^{3 \tau} D_B \tau \left[ D_A D_C C^{BC}(e^\tau
      u,\theta^A) - D^B D^C C_{CA}(e^\tau u,\theta^A) \right]
\nonumber \\ &&
+ \frac{3}{4} u e^{3 \tau} D_B \tau \, C_{AC}(e^\tau u,\theta^A) \,
N^{BC}(e^\tau u,\theta^A).
  \end{eqnarray}
\end{subequations}

To derive the transformation laws (\ref{wnonlinst}), we again extend the
diffeomorphism (\ref{weyl1}) to a diffeomorphism $\eta : M \to M$ which
coincides with $\psi$ on ${\mathscr I}^+$, by
assuming a general power series expansion in $1/r$ and 
using the notation $x^\alpha =
(u,r,\theta^A)$ and ${\bar x}^\alpha = x^\alpha \circ \eta$:
\begin{subequations}
  \label{nonlinst3}
  \begin{eqnarray}
    {\bar u} &=& e^{\tau(\theta^A)} \left[ u  + \frac{1}{r}
    \beta^{(1)}(u,\theta^A)
    + \frac{1}{r^2} \beta^{(2)}(u,\theta^A)\right] + O\left( \frac{1}{r^3} \right),\\
{\bar r} &=& e^{-\tau(\theta^A)} \left[ r + R^{(1)}(u,\theta^A) + \frac{1}{r} R^{(2)}(u,\theta^A)\right]
+ O\left( \frac{1}{r^2} \right),\\
\label{nonlinst3c}
{\bar \theta}^A &=& \theta^A + \frac{1}{r} \chi^{(1)A}(u,\theta^A)
+ \frac{1}{r^2} {\hat \chi}^{(2)A}(u,\theta^A)
+ \frac{1}{r^3} {\hat \chi}^{(3)A}(u,\theta^A)
+ O\left( \frac{1}{r^4} \right).
  \end{eqnarray}
\end{subequations}
Here as before the functions $\beta^{(1)}$, $\beta^{(2)}$, $R^{(1)}$, $R^{(2)}$,
$\chi^{(1)}$, ${\hat \chi}^{(2)}$ and ${\hat \chi}^{(3)}$ are
arbitrary, and we use instead of
${\hat \chi}^{(2)A}$ and ${\hat \chi}^{(3)A}$ the covariant quantities
$\chi^{(2)A}$ and $\chi^{(3)A}$ defined by Eqs.\ (\ref{hatchif}).

We next take the pullback ${\bar g}_{ab} = \eta_* g_{ab}$ of the
metric and follow the same steps as in Sec.\ \ref{sec:sts}.
From $O(rA,0)$, the trace part of $O(AB,-1)$,
and $O(rr,2)$, 
respectively, we find
that
\begin{subequations}
  \begin{eqnarray}
    \chi^{(1)A} &=& - u e^{2 \tau} h^{AB} D_B \tau, \\
    R^{(1)}& =& \frac{1}{2} u e^{2 \tau} h^{AB} (D_A\tau D_B\tau + D_A D_B \tau), \\
    \beta^{(1)} &=& - \frac{1}{2} u^2 e^{2 \tau}  (D \tau)^2. 
  \end{eqnarray}
  \end{subequations}
We then obtain from $O(AB,-2)$ and from the trace-free part of
$O(AB,-1)$ the transformation 
laws (\ref{wnonlinsta0}) and (\ref{wnonlinsta})
for the metric $h_{AB}$ and shear tensor $C_{AB}$.

Next we find from $O(rA,1)$, $O(rr,3)$, $O(ru,2)$ and $O(rA,2)$
respectively that
\begin{subequations}
  \begin{eqnarray}
    \chi^{(2)A} & =& \frac{1}{2} u e^{3 \tau} C^{AB}(e^\tau u,\theta^A)
    D_B \tau
\nonumber \\ &&
    + \frac{1}{2} u^2 e^{4 \tau} \left( D^A\tau D^2 \tau -
      D^A D_B \tau D^B \tau - D^A \tau D_B\tau D^B \tau \right), \\
\beta^{(2)}& =& \frac{1}{4} u^2 e^{3 \tau} C_{AB}(e^\tau u,\theta^A) D^A \tau D^B
\tau + \frac{1}{4} u^3 e^{3 \tau} (D \tau)^2 D^2 e^\tau, \\
R^{(2)} &=&
- \frac{1}{4} u e^{2 \tau} C^{AB}(e^\tau u,\theta^A) D_A D_B e^\tau
- \frac{1}{4} u^2 e^{4 \tau} N^{AB}(e^\tau u,\theta^A) D_A\tau D_B
\tau
\nonumber \\ &&
- \frac{1}{2} e^{3 \tau} u D_A C^{AB}(e^\tau u,\theta^A) D_B \tau
+ \frac{1}{4} e^{2 \tau} u^2 \left[ D_A D_B e^\tau D^A
D^B e^\tau - \frac{1}{2} (D^2 e^\tau)^2 \right]
\nonumber\\ && 
+ \frac{1}{4} e^{4 \tau} u^2 \ricci (D \tau)^2, \\
\label{chi3c}
    \chi^{(3)A} &=& - \frac{1}{16} u e^{4 \tau} D^A \tau C_{BC}(e^\tau u,\theta^A)
    C^{BC}(e^\tau u,\theta^A)
+ \frac{1}{4} u^2 e^{5 \tau} D_B C^{AC}(e^\tau u,\theta^A) D^B\tau D_C\tau
\nonumber \\ &&
- \frac{1}{4} u^2 e^{5 \tau} D^B C_{BC}(e^\tau u,\theta^A) D^A\tau D^C\tau
+ \frac{1}{4} u^2 e^{4 \tau} C_{BC}(e^\tau u,\theta^A) D^A\tau D^B D^Ce^\tau
    \nonumber \\
    &&    - \frac{1}{2} u^2 e^{4 \tau} C_{BC}(e^\tau u,\theta^A) D^C\tau D^A D^Be^\tau
+ u^2 e^{5 \tau} C^{AB}(e^\tau u,\theta^A) D_B\tau (D \tau)^2
 \nonumber \\
&&+     \frac{1}{3} u^3 e^{6 \tau} N^{AB}(e^\tau u,\theta^A) D_B \tau (D \tau)^2     - \frac{1}{6} u^3 e^{6\tau} N^{BC}(e^\tau u,\theta^A) D^A\tau D_B\tau D_C \tau
    \nonumber \\ 
&&  +       \frac{1}{2} u^3 e^{5 \tau} D^A\tau D^B \tau D_B D^2e^\tau 
    - \frac{1}{3} u^3 e^{5 \tau} D^B\tau D^C \tau D^A D_B D_C e^\tau      \nonumber \\
    &&      - \frac{5}{24} u^3 e^{6 \tau} D^A\tau (D^2 \tau)^2
    + \frac{1}{12} u^3 e^{6 \tau} D^A\tau D_B D_C \tau D^B D^C \tau
    \nonumber \\ &&
    + \frac{1}{6} u^3 e^{6\tau} D_B \tau D^A D^B \tau  D^2 \tau
- \frac{5}{6} u^3 e^{6 \tau} D^A D^B \tau D_B \tau (D \tau)^2
+ \frac{3}{4} u^3 e^{6 \tau} D^A \tau D^2 \tau (D \tau)^2
\nonumber \\
&& -\frac{1}{2} u^3 e^{6 \tau} D^A \tau D_B D_C \tau D^B \tau D^C \tau
- \frac{23}{24} u^3 e^{6 \tau} D^A \tau (D\tau)^4
\nonumber \\ &&
+ \frac{1}{6} u^3 e^{ 6\tau} D^A \tau (D\tau)^2 \ricci. 
  \end{eqnarray}
  \end{subequations}
Here the angular indices $A$, $B$ are raised and lowered with $h_{AB}$
and not ${\bar h}_{AB}$ given by Eq.\ (\ref{wnonlinsta0}).
Finally from $O(uu,1)$ and $(uA,1)$ we obtain the transformation laws
(\ref{wnonlinstb}) and (\ref{wnonlinstc}) for the Bondi mass aspect $m$ and angular
momentum aspect $N_A$.

\subsection{Two-sphere diffeomorphisms of the Weyl BMS group}
\label{sec:diffs2}

The third category of transformations in the BMSW group are
diffeomorphisms of the two sphere:
\be
   {\bar u} = u, \ \ \ \ {\bar \theta}^A = \chi^A(\theta^B).
   \label{2sd}
\ee
This extends to the exact four-dimensional diffeomorphism
\be
{\bar u} = u, \ \ \ \ {\bar \theta}^A = \chi^A(\theta^B), \ \ \ \ {\bar r} = r,
\ee
under which the metric functions transform by the pullback of the two-sphere diffeomorphism $\chi$:
\begin{subequations}
\label{anonlinst}
\begin{eqnarray}
  \label{anonlinsta0}
  {\bar h}_{AB} &=& \chi_* h_{AB},  \\
  \label{anonlinsta}
    {\bar C}_{AB} &=& \chi_* C_{AB}, \\
  \label{anonlinstb}    
        {\bar m} &=& \chi_* m, \\
      \label{anonlinstc}
    {\bar N}_A &=& \chi_* N_A.
  \end{eqnarray}
\end{subequations}

\subsection{Two-sphere diffeomorphisms of the generalized BMS group}
\label{2sd:gbms}

We now turn to the generalized BMS group instead of the Weyl BMS group,
and focus again on two-sphere diffeomorphisms.  The diffeomorphism
$\psi$ on ${\mathscr I}^+$ takes the form
\be
\label{gbms1}
   {\bar u} = e^{\alpha(\theta^A)} u, \ \ \ \ {\bar \theta}^A = \chi^A(\theta^B).
\ee
Here the function $\alpha(\theta^A)$ is determined as a function of $\chi$
by the requirement (\ref{tauval}).

To compute the transformation of the metric functions under the mapping
(\ref{gbms1}) we combine the results of Secs.\ \ref{sec:conf} and
\ref{sec:diffs2}. 
We decompose $\psi$ as $\psi = \psi_2 \circ \psi_1$, where $\psi_1$ is
the two-sphere diffeomorphism (\ref{2sd}),
and $\psi_2$ is the conformal transformation (\ref{weyl1}) with $\tau$
chosen to be 
\be
\tau = \alpha \circ \chi^{-1}.
\label{tauval1}
\ee
The corresponding spacetime diffeomorphisms are related by $\eta =
\eta_2 \circ \eta_1$, and the pullback of the metric is then given by
\be
\eta_* g_{ab} = \eta_{1\,*} \circ \eta_{2\,*} g_{ab}.
\ee
It follows that we can compute transformed metric functions as
follows.
Start with the metric functions $h_{AB}, m, N_A, C_{AB}$, and 
act with the pullback $\eta_{2\,*}$ which yields the transformed
metric functions ${\bar h}_{AB}, {\bar m}, {\bar N}_A, 
{\bar C}_{AB}$ given by Eqs.\ (\ref{wnonlinst}) using the parameter (\ref{tauval1}).  Next, 
act with the pullback $\eta_{1\,*}$ using the prescription
(\ref{anonlinst}) which yields the final metric functions
\be
\label{gg}
{\hat h}_{AB} = \chi_* {\bar h}_{AB}, \ \ \
   {\hat m} = \chi_* {\bar m}, \ \ \
   {\hat N}_A = \chi_* {\bar N}_A, \ \ \ 
   {\hat C}_{AB} = \chi_* {\bar C}_{AB}.
   \ee
In particular from Eq.\ (\ref{wnonlinsta0}) the final induced metric
is given by 
$
   {\hat h}_{AB} = \chi_* (e^{-2 \tau}  h_{AB}),
   $
   and so the final volume form is 
   \be
   \label{finalep}
   {\hat \epsilon}_{AB} = \chi_*( e^{-2 \tau } \epsilon_{AB})  = e^{-2
     \alpha} \chi_* \epsilon_{AB} = \frac{ e^{-2 \alpha}
   }{\omega_\chi} \epsilon_{AB} = \pm \epsilon_{AB},
   \ee
where we have used Eqs.\ (\ref{tauval1}), (\ref{not:wdef}) and
(\ref{tauval}).

\subsection{Conformal isometries of the BMS group}

We now turn to the boosts and rotations of the BMS group.
These are a special case of the GBMS two-sphere diffeomorphisms 
(\ref{gbms1}) where $\chi$ is restricted to be a global conformal isometry,
so that
\be
\chi_* h_{AB} = \frac{1}{\omega_\chi} h_{AB}.
\ee
Here $\omega_\chi$ is defined by Eq.\ (\ref{not:wdef}), and we
are excluding improper Lorentz transformations.
The transformation laws are therefore given by combining
Eqs.\ (\ref{wnonlinst}), (\ref{tauval1}) and (\ref{gg}).
It follows from the
analysis that led to Eq.\ (\ref{finalep})  that the two-metric is preserved,
$
   {\hat h}_{AB} = h_{AB}.
$
Moreover for rotations we have $\omega_\chi=1$, and for boosts we have
 $
 \omega_\chi = (\cosh \gamma - \cos \Theta \sinh \gamma)^2,
 $
 where $\gamma$ is the rapidity parameter and $\Theta$ is the angle between the boost direction and the direction determined by $\theta^A$.
 From Eqs.\ (\ref{not:wdef}), (\ref{tauval}) and (\ref{tauval1}) we obtain that
 \be
\label{simple}
 e^{2 \tau} = \frac{1}{\omega_\chi \circ \chi^{-1}} =
 \omega_{\chi^{-1}} = (\cosh \gamma + \, \cos \Theta \, \sinh
 \gamma)^2.
 \ee
It follows that $e^{\tau}$ is purely $l=0$ and $l=1$, and so it is
annihilated by the differential operator $D_A D_B - h_{AB} D^2/2$.
This implies that the following terms vanish in
Eqs.\ (\ref{wnonlinst}) for boosts and rotations:
the second term in Eq.\ (\ref{wnonlinsta}), and the second, third and
sixth terms in Eq.\ (\ref{wnonlinstb}).  In addition the terms
involving $\ricci$ in Eqs.\ (\ref{nonlinst}) and
(\ref{wnonlinst}) will vanish when we specialize to round two-metrics with
$\ricci = 2$, as is normally done for the BMS group.

\section{Comparisons with previous results} \label{sec:comparison}

Our results
(\ref{nonlinst}), (\ref{wnonlinst}), (\ref{anonlinst}) and (\ref{gg})
agree (mostly) with a number of previous computations in special cases:

\begin{itemize}

\item The linearized BMSW computations of Freidel, Oliveri, Pranzetti
  and Speziale given in Eqs.\  (4.42) of Ref.\ \cite{Freidel:2021fxf},
  taking into account that the variables $({\bar F},T,W,\tau,{\bar
    P}_A)$ used there are given in terms of variables used here to
  linear order
  as $(\ricci/4,\beta,\tau, \beta + u \tau, N_A - D_A (C_{BC}
  C^{BC})/16 - C_{AB} D_C C^{BC}/4)$.

\item The linearized GBMS computations of
Comp\`ere, Fiorucci and Ruzziconi given in Eqs.\ (2.20) -- (2.24) of
Ref.\ \cite{Compere:2018ylh}, noting that their angular momentum
aspect $N_A$ is related to ours by their Eq.\ (2.8).

\item Our nonlinear computations in the BMS context given in Appendix B of Ref.\ \cite{FN}
  and Appendix B of Ref.\ \cite{Flanagan:2021svq}.  
    These method used in those computations was to combine the
    Bondi-coordinate charge expression (FN,3.5) of
  Ref.\ \cite{FN}, which is known to be 
  covariant, together with known nonlinear transformation of the
  symmetry generator vector fields on $\scri^+$ to indirectly deduce
  the transformations of the metric functions.  The results are limited
  to nonradiative regimes where $N_{AB}=0$ since the charge expression
  (FN,3.5) was derived only in that context.
  Demonstrating consistency with the results of this paper
requires using (i) the condition $N_{AB}=0$; (ii) the simplifications discussed after
  Eq.\ (\ref{simple}) above; (iii) the definition (3.6) of 
  Ref.\ \cite{Flanagan:2021svq} of the variable ${\hat N}_A$; and (iv)
  the evolution equation (FN,2.11a) for the Bondi mass aspect which
  shows that ${\dot m}=0$ in vacuum when $N_{AB}=0$.

  \item The nonlinear BMS transformation laws derived in Appendices C.5 and
    C.6 of the book \cite{2002hftr.book.....C} by Chrusciel,
    Jezierski, and Kijowski, using the fact that their angular
    momentum aspect is related to ours by a factor of $-3$.  For
    supertranslations the results agree, except for the transformation law for
    $N_A$, their Eq.\ (C.124).  The difference seems to arise from
    a discrepancy between our Eq.\ (\ref{beta2ans}) for the function
    $\beta^{(2)}$ compared to their corresponding Eq.\ (C.114).
    For boosts the results do not agree; we have not been able to
    track down the source of the discrepancy in this case.

  \item The nonlinear BMS transformation laws recently derived by
Chen, Wang, Wang and Yau in Ref.\ \cite{Chen:2023zpl}, which are
consistent with our results when the simplifications discussed after
Eq.\ (\ref{simple}) above are used.

\end{itemize}

\section{Applications} \label{sec:applications}

\subsection{Vacuum structure}

One application of our nonlinear transformation results 
(\ref{nonlinst}), (\ref{wnonlinst}), (\ref{anonlinst}) and (\ref{gg})
is to obtain an explicit parameterization of ``vacuum'' states which
in a local region of $\scri^+$ are diffeomorphic to the data for
Minkowski spacetime.  This allows us to reproduce and generalize
slightly the GBMS results of Comp\`ere, Fiorucci and Ruzziconi given
in Sec.\ 3 of Ref.\ \cite{Compere:2018ylh}.

The result is as follows.  First, following
Refs.\ \cite{Compere:2018ylh,Freidel:2021fxf,Freidel:2021qpz}
we define the Liouville or Geroch tensor $N_{AB}^{\rm vac}[h_{CD}]$, a
function of the metric $h_{CD}$, as follows.
We choose a conformal factor $e^{2 \tau}$ to make $e^{2 \tau} h_{AB}$
a unit round two-metric, by solving
the equation
\be
 2 D^2 \tau + 2 e^{2 \tau} = {\mathcal R}
\label{taueqn}
\ee
for $\tau$.  We then define the tensor
\be
\label{geroch0}
N_{AB}^{\rm vac} = 2 e^{\tau} (D_A D_B - 
h_{AB} D^2/2) e^{-\tau},
\ee
which
has the property
\be
\label{geroch}
D^A N_{AB}^{\rm vac} = - \frac{1}{2} D_B {\mathcal R}.
\ee
General vacuum data can now be parameterized in terms of an arbitrary
two-metric $h_{AB}(\theta^C)$ and a function $\beta(\theta^C)$ as follows:
\begin{subequations}
  \label{final}
  \begin{eqnarray}
    \label{NABvac}
    N_{AB} &=& N_{AB}^{\rm vac}[h_{CD}], \\
    \label{finalshear}
  C_{AB} &=& (u + \beta) N^{\rm vac}_{AB}[h_{CD}] - 2 (D_A D_B -   h_{AB} D^2/2) \beta, \\
m &=& - \frac{1}{8}   C^{AB} N_{AB}^{\rm vac}[h_{CD}], \\
N_A &=& 0.
  \end{eqnarray}
  \end{subequations}
The result (\ref{final}), when restricted to metrics $h_{AB}$ whose
volume form is fixed as appropriate for the GBMS group,
agrees with Eqs.\ (3.10) and (3.26) of
\cite{Compere:2018ylh} \footnote{The comparison uses
  their relation (2.8) between the two different angular momentum aspects, and
  that their variables
$q_{AB}$, $\Phi$ and $C$ are 
  given in terms of ours by $q_{AB} = h_{AB}$, $C = \beta$ and $\Phi = 2 \tau - \ln \gamma_s$.
  Here $\gamma_s$ is chosen so that the metric $\gamma_s^{-1} e^{2 \tau} h_{AB}$ is flat.
  It is given explicitly by $\gamma_s = 2 / (1 + z {\bar z})^2$, where $(z,{\bar z})$ 
 are the complex stereographic coordinates associated with the round metric $e^{2 \tau} h_{AB}$.}.
The result (\ref{final}) is also consistent with the results of Sec.\ 4.7 of
Ref.\ \cite{Freidel:2021fxf} on the vacuum structure.

We obtain the form
(\ref{final}) of general vacuum data by applying a general BMSW
transformation to Minkowski data.
We parameterize the BMSW transformation by composing the two
sphere diffeomorphism (\ref{2sd}), the conformal transformation
(\ref{weyl1}), and the supertranslation (\ref{st1}).
We start with the Minkowski data
\be
\mathring{h}_{AB}, \ \ \ \ {\mathring C}_{AB}=0, \ \ \ \ \mathring{m}
= 0, \ \ \ \ \mathring{N}_A =0,
\ee
where $\mathring{h}_{AB}$ is a unit round two metric.  Acting with the
two-sphere diffeomorphism $\chi$ using Eq.\ (\ref{anonlinst}) gives the data
\be
{\bar h}_{AB} = \chi_* \mathring{h}_{AB}, \ \ \ \ {\bar C}_{AB}=0,
\ \ \ \ {\bar m}
= 0, \ \ \ \ {\bar N}_A =0.
\ee
Next acting with the conformal transformation $\tau$ using
Eqs.\ (\ref{wnonlinst}) gives
\begin{subequations}
  \begin{eqnarray}
    \label{hath}
    {\hat h}_{AB} &=& e^{-2 \tau} \chi_* \mathring{h}_{AB}, \\
    {\hat N}_{AB} &=& - 2 e^{-\tau} ({\bar D}_A {\bar D}_B - {\bar
      h}_{AB} {\bar D}^2/2) e^\tau
= 2 e^{\tau} ({\hat D}_A {\hat D}_B - {\hat
  h}_{AB} {\hat D}^2/2) e^{-\tau}, \\
    {\hat C}_{AB} &=& u {\hat N}_{AB}, \\
{\hat m} &=& - \frac{1}{8} u {\hat N}_{AB} {\hat N}_{CD} {\hat h}^{AC}
{\hat h}^{BD}, \\
{\hat N}_A &=& 0.
  \end{eqnarray}
  \end{subequations}
Here ${\bar D}_A$ and ${\hat D}_A$ are the derivative operators
associated with ${\bar h}_{AB}$ and ${\hat h}_{AB}$ and we used ${\bar
  {\mathcal R}} =2$.
From Eq.\ (\ref{hath}) we can obtain generic two metrics ${\hat
  h}_{AB}$ by choosing $\chi$ and $\tau$ appropriately, and it will be
convenient following \cite{Compere:2018ylh} to use ${\hat
    h}_{AB}$ to parameterize the state rather than $\chi$ and $\tau$. 
In the GBMS context of \cite{Compere:2018ylh} the conformal
transformation $\tau$ is constrained to be a function of $\chi$, which
constrains ${\hat h}_{AB}$ to have the same volume form as
$\mathring{h}_{AB}$, as discussed in Sec.\ \ref{2sd:gbms} above.

The final step is to act with the supertranslation $\beta$ using
Eqs.\ (\ref{nonlinst}).
This yields
\begin{subequations}
  \label{xfinal}
  \begin{eqnarray}
    \label{xfinal0}
    h_{AB} &=& {\hat h}_{AB}, \\
    \label{xNABvac}
    N_{AB} &=& {\hat N}_{AB} = 2 e^{\tau} (D_A D_B - 
    h_{AB} D^2/2) e^{-\tau}, \\
    \label{xfinalshear}
  C_{AB} &=& (u + \beta) N_{AB} - 2 (D_A D_B -   h_{AB} D^2/2) \beta, \\
m &=& - \frac{1}{8}   C_{AB} N_{CD} h^{AC}
h^{BD}, \\
N_A &=& 0.
  \end{eqnarray}
  \end{subequations}
Here Eqs.\ (\ref{xfinal0}) and (\ref{hath})
yield the relation (\ref{taueqn})
that determines $\tau$ in terms of the final metric $h_{AB}$.
We used the identities (\ref{geroch}) and 
${\hat N}_{AB} {\hat N}^B_{\ \,C} = {\hat N}_{BD}
{\hat N}^{BD} {\hat h}_{AC}/2$, which is valid for any symmetric traceless
tensor in two dimensions.
The results (\ref{xfinal}) together with the definition
(\ref{geroch0}) now yield the parameterization (\ref{final}).

\subsection{Other coordinates on phase space}

The choice of Bondi coordinate metric functions is somewhat arbitrary,
and is useful to consider other coordinates on the phase space that
simplify the description.  One organizing principle is to use
variables which vanish in vacuum regions of $\scri^+$
\cite{Compere:2016jwb,Compere:2018ylh}. Another is to use variables
with simple transformation properties under conformal transformations
\cite{Freidel:2021fxf,Freidel:2021qpz,Freidel:2021dfs,Freidel:2021ytz,Pasterski:2017kqt,Raclariu:2021zjz}.
These considerations lead to the following modified definitions of
Bondi mass aspect, shear, and news tensor
\cite{Compere:2018ylh,Freidel:2021fxf,Freidel:2021qpz}: 
\begin{subequations}
  \label{newvars}
  \begin{eqnarray}
    {\mathcal M} &=& m +\frac{1}{8}  C_{AB} N^{AB}
    , \\
    {\mathcal C}_{AB} &=& C_{AB} - u N_{AB}^{\rm vac}[h_{CD}], \\
    {\mathcal N}_{AB} &=& \partial_u {\cal C}_{AB} = N_{AB} - N_{AB}^{\rm vac}[h_{CD}],
  \end{eqnarray}
\end{subequations}
where $N_{AB}^{\rm vac}$ is defined in Eq.\ (\ref{geroch0}).
We do not need to redefine the angular momentum aspect
$N_A$ \footnote{Note that our definition of $N_A$ coincides with the
covariant angular momentum aspect ${\cal P}_A$ of Refs.\ \cite{Freidel:2021fxf,Freidel:2021qpz}.}.
With these definitions ${\cal M}$, ${\cal N}_{AB}$ and $N_A$ vanish in
vacuum regions 
from Eqs.\ (\ref{final}) \footnote{Conversely, the condition that
${\cal M}, {\cal N}_{AB}$ and $N_A$ all vanish is not sufficient to imply
that the region is of the vacuum form (\ref{final}).  The additional
condition needed is that 
the shear $C_{AB}$ must have the form (\ref{finalshear}), which is
equivalent 
to the 
vanishing of the dual covariant mass (\ref{dcm}) given that ${\mathcal N}_{AB}=0$.}.	

We can rewrite our transformation laws 
(\ref{nonlinst}) and (\ref{wnonlinst}) in terms of the new variables (\ref{newvars}),
extending the linearized transformation laws of Secs. 4.5 of
Ref.\ \cite{Freidel:2021fxf} and 2.2 of Ref.\ \cite{Freidel:2021qpz}.
Under supertranslations we find from Eqs.\ (\ref{nonlinst}), (\ref{newvars}) and
(\ref{geroch}) that
\begin{subequations}
\label{nonlinstnew}
\begin{eqnarray}
  \label{nonlinstanew}
        {\bar h}_{AB}(\theta^C) &=& h_{AB}(\theta^C), \\
  \label{nonlinstanew1}        
    {\bar {\mathcal C}}_{AB}(u,\theta^A) &=& {\mathcal C}_{AB}(u + \beta, \theta^A) - 2 D_A D_B \beta
    + h_{AB} D^2 \beta + \beta N_{AB}^{\rm vac}[h_{CD}],\\
        {\bar {\mathcal N}}_{AB}(u,\theta^A) &=& {\mathcal N}_{AB}(u + \beta, \theta^A),\\
  \label{nonlinstbnew}    
    {\bar {\mathcal M}}(u,\theta^A) &=& {\mathcal M}(u + \beta,
    \theta^A)
 + \frac{1}{2} D_A {\mathcal N}^{AB}(u+\beta,\theta^A) D_B\beta
    \nonumber \\     && 
     +  \frac{1}{4} {\dot {\mathcal N}}^{AB}(u+\beta,\theta^A) D_A \beta
     D_B\beta,\\
\label{nonlinstcnew}
     {\bar N}_A(u,\theta^A) &=& N_A(u+\beta,\theta^A) + 3
    {\mathcal M}(u+\beta,\theta^A)  D_A\beta   
    \nonumber
+ 3 {\mathscr M}(u+\beta,\theta^A) \epsilon_A^{\ \,B} D_B \beta
    \\ &&
+ \frac{3}{2} D_A\beta \,D_B\beta \, D_C {\mathcal N}^{BC}(u+\beta,\theta^A)
- \frac{3}{4} (D \beta)^2 \,D^C {\mathcal N}_{AC}(u+\beta,\theta^A)
\nonumber \\ &&
+ \frac{1}{2} D_A\beta \, D_B \beta \, D_C \beta {\dot {\mathcal N}}^{BC}(u+\beta,\theta^A)
\nonumber \\ &&
- \frac{1}{4} (D\beta)^2 \, D^C \beta {\dot  {\mathcal N}}_{CA}(u+\beta,\theta^A). \ \ \ \ \ \ \ \
  \end{eqnarray}
\end{subequations}
Here ${\mathscr M}$ is the dual covariant mass defined by Freidel and
Pranzetti \cite{Freidel:2021qpz} given by
\be
\label{dcm}
   {\mathscr M} = \frac{1}{8} C_{AB} \epsilon^{BC} N^A_{\ \, C}
+ \frac{1}{4} D_A D^B \epsilon^{AC} C_{BC},
\ee
and the terms linear in $\beta$ in Eq.\ (\ref{nonlinstcnew}) have
  been manipulated as described there.
Note that the transformation laws (\ref{nonlinstnew}) 
are much simpler than the original versions
(\ref{nonlinst}).

For conformal transformations we first compute the transformation
law for $N_{AB}^{\rm vac}[h_{CD}]$.  Denoting the solution of
Eq.\ (\ref{taueqn}) by $\tau[h_{CD}]$, we find that
$
\tau[e^{2 \psi} h_{CD}] = \tau[h_{CD}] - \psi,
$
and it follows from the definition (\ref{geroch0}) that [cf.\ Eq.\ (4.56) of
  Ref.\ \cite{Freidel:2021fxf}] 
\be
N_{AB}^{\rm vac}[e^{2 \psi} h_{CD}] = N_{AB}^{\rm vac}[h_{CD}] -2
e^\psi (D_A D_B -  h_{AB} D^2/2) e^{-\psi}.
\ee
Now from Eqs.\ (\ref{wnonlinst}),
(\ref{newvars}) and (\ref{geroch}) we find that
\begin{subequations}
\label{wnonlinstnew}
\begin{eqnarray}
  \label{wnonlinsta0new}
  {\bar h}_{AB}(\theta^A) &=& e^{-2 \tau} h_{AB}(\theta^A),  \\
  \label{wnonlinstanew}
    {\bar {\mathcal C}}_{AB}(u,\theta^A) &=& e^{-\tau} {\mathcal
      C}_{AB}(e^\tau u, \theta^A),\\
        {\bar {\mathcal N}}_{AB}(u,\theta^A) &=& {\mathcal
          N}_{AB}(e^\tau u, \theta^A),\\
  \label{wnonlinstbnew}    
        {\bar {\mathcal M}}(u,\theta^A) &=& e^{3 \tau} {\mathcal M}(e^\tau u, \theta^A)
        + \frac{1}{2} u e^{4 \tau} D_A {\mathcal N}^{AB}(e^\tau u,\theta^A) D_B\tau
        \nonumber \\
        &&
+ \frac{1}{4} e^{5 \tau} u^2 {\dot {\mathcal N}}^{AB}(e^\tau u, \theta^A) D_A\tau
D_B \tau, \\
\label{wnonlinstcnew}
     {\bar N}_A(u,\theta^A) &=& e^{2 \tau} N_A(e^\tau u,\theta^A) + 3
    u e^{3 \tau} {\mathcal M}(e^\tau u,\theta^A)  D_A\tau   
    \nonumber
+ 3 u e^{3 \tau} {\mathscr M}(e^\tau u,\theta^A) \epsilon_A^{\ \,B} D_B \tau
    \\ &&
+ \frac{3}{2} u^2 e^{4 \tau} D_A\tau \,D_B\tau \, D_C {\mathcal
  N}^{BC}(e^\tau u,\theta^A)
- \frac{3}{4} u^2 e^{4 \tau} (D \tau)^2 \,D^C {\mathcal N}_{AC}(e^\tau u,\theta^A)
\nonumber \\ &&
+ \frac{1}{2} u^3 e^{5 \tau} D_A\tau \, D_B \tau \, D_C \tau {\dot
  {\mathcal N}}^{BC}(e^\tau u,\theta^A)
\nonumber \\ &&
- \frac{1}{4} u^3 e^{5 \tau} (D \tau)^2 \, D^C \tau {\dot {\mathcal N}}_{CA}(e^\tau u,\theta^A). \ \ \ \ \ \ \ \
  \end{eqnarray}
\end{subequations}
Again these transformation laws are much simpler than the original
versions (\ref{wnonlinst}).

From Eqs.\ (\ref{nonlinstnew}) and (\ref{wnonlinstnew}) it follows that the transformation
laws for the dual covariant mass ${\mathscr M}$ have the same forms
(\ref{nonlinstbnew}) and (\ref{wnonlinstbnew}) as for the mass
${\mathcal M}$,
except that the modified news tensor ${\mathcal N}_{AB}$ is replaced
by its dual ${\tilde {\mathcal N}}_{AB} = \epsilon_{AC} {\mathcal
  N}^C_{\ \,B}$.  This was previously shown to linear order in Ref.\ \cite{Freidel:2021qpz}. 

We also note that a different kind of simplification can achieved by defining
the modified mass aspect
\be
\mathbb{M} =  m - \frac{1}{4} D^A D^B C_{AB}.
\ee
The transformation laws for this quantity under supertranslations and conformal transformations are
\be
   {\bar {\mathbb{M}}}(u,\theta^A) =  \mathbb{M}(u + \beta,\theta^A) + \frac{1}{4} D^4 \beta + \frac{1}{4} \mathcal{R} D^2 \beta
+ \frac{1}{2} D^A{\mathcal R} D_A\beta,
   \ee
   and
  \begin{eqnarray}
    \label{jkjk}
   {\bar {\mathbb{M}}}(u,\theta^A) &=&  e^{3\tau} \mathbb{M}(e^\tau u ,\theta^A) + \frac{1}{4} u e^{3 \tau} \left[ D^4 e^\tau +  \mathcal{R} D^2 e^\tau
     + 2 D^A{\mathcal R} D_Ae^\tau \right] \nonumber \\
   && + \frac{1}{2} e^{2 \tau} C^{AB}(e^\tau u,\theta^A) D_A D_B e^\tau
   - u e^{3\tau} (D^A D^B e^\tau) (D_A D_B - h_{AB} D^2/2) e^\tau  .\ \ \ \ \ \ \ \ \ 
\end{eqnarray}
The terms on the right hand side in Eq.\ (\ref{jkjk}) other than the first term vanish in the BMS context,
when ${\mathcal R} =2$ and $e^\tau$ is purely $l=0$ and $l=1$.
This mass definition is therefore natural to use in the BMS context,
but somewhat less so in the more general BMSW context. 

\acknowledgments

We thank Bob Wald for a helpful conversation and for pointing out to us
the reference \cite{Chen:2023zpl}.


\bibliographystyle{JHEP}
\bibliography{infoloss,asycps}

\end{document}